\documentclass[letterpaper,twocolumn,prl,aps,superscriptaddress,amsmath,amssymb,floatfix]{revtex4-1}
\usepackage{mathptmx}
\usepackage[latin9]{inputenc}
\usepackage{color}
\usepackage{amsmath}
\usepackage{amssymb}
\usepackage{graphicx}
\usepackage{esint}
\usepackage{bm}
\usepackage{enumitem}
\usepackage{ulem}
\usepackage[unicode=true,
bookmarks=true,bookmarksnumbered=false,bookmarksopen=false,
 breaklinks=false,pdfborder={0 0 1},backref=false,colorlinks=true]
 {hyperref}
\hypersetup{
 linkcolor=magenta,urlcolor=blue,citecolor=blue,pdfstartview={FitH}}

\makeatletter

\pdfpageheight\paperheight
\pdfpagewidth\paperwidth

\usepackage{textcomp}
\usepackage{epstopdf}

\pdfpageheight\paperheight
\pdfpagewidth\paperwidth

\@ifundefined{textcolor}{}{%
 \definecolor{BLACK}{gray}{0}
 \definecolor{WHITE}{gray}{1}
 \definecolor{RED}{rgb}{1,0,0}
 \definecolor{GREEN}{rgb}{0,1,0}
 \definecolor{BLUE}{rgb}{0,0,1}
 \definecolor{CYAN}{cmyk}{1,0,0,0}
 \definecolor{MAGENTA}{cmyk}{0,1,0,0}
 \definecolor{YELLOW}{cmyk}{0,0,1,0}
}

\usepackage{xcolor}
\usepackage{soul}
\newcommand{\bra}[1]{\ensuremath{\left\langle#1\right|}}
\newcommand{\ket}[1]{\ensuremath{\left|#1\right\rangle}}

\definecolor{blue}{rgb}{0,0,1}
\definecolor{red}{rgb}{1,0,0}
\definecolor{green}{rgb}{0,1,0}

\usepackage{soul}

\makeatother

\begin{document}
\title{Scalable Optical Links for Controlling Bosonic Quantum Processors}

\author{Chuanlong~Ma}
\thanks{These authors contributed equally to this work.}
\affiliation{Center for Quantum Information, Institute for Interdisciplinary Information
Sciences, Tsinghua University, Beijing 100084, China}

\author{Jia-Qi~Wang}
\thanks{These authors contributed equally to this work.}
\affiliation{Laboratory of Quantum Information, University of Science and
Technology of China, Hefei 230026, China.}

%ÀîÁÖÔó£¬ÍõÂ³Óñ£¬³ÂÇïÊµ
\author{Linze~Li}
\thanks{These authors contributed equally to this work.}
\affiliation{School of Information Science and Technology, ShanghaiTech University, Shanghai, China}

\author{Jiajun~Chen}
\thanks{These authors contributed equally to this work.}
\affiliation{Center for Quantum Information, Institute for Interdisciplinary Information
Sciences, Tsinghua University, Beijing 100084, China}

\author{Xiaoxuan~Pan}
\affiliation{Center for Quantum Information, Institute for Interdisciplinary Information
Sciences, Tsinghua University, Beijing 100084, China}

\author{Zheng-Hui~Tian}
\affiliation{Laboratory of Quantum Information, University of Science and
Technology of China, Hefei 230026, China.}

\author{Zheng-Xu~Zhu}
\affiliation{Laboratory of Quantum Information, University of Science and
Technology of China, Hefei 230026, China.}

\author{Jia-Hua~Zou}
\affiliation{Laboratory of Quantum Information, University of Science and
Technology of China, Hefei 230026, China.}

\author{Dingran~Gu}
\affiliation{School of Information Science and Technology, ShanghaiTech University, Shanghai, China}

\author{Luyu~Wang}
\affiliation{School of Information Science and Technology, ShanghaiTech University, Shanghai, China}

\author{Qiushi~Chen}
\affiliation{School of Information Science and Technology, ShanghaiTech University, Shanghai, China}

\author{Weiting~Wang}
\affiliation{Center for Quantum Information, Institute for Interdisciplinary Information
Sciences, Tsinghua University, Beijing 100084, China}

\author{Xin-Biao~Xu}
\affiliation{Laboratory of Quantum Information, University of Science and
Technology of China, Hefei 230026, China.}
\affiliation{Anhui Province Key Laboratory of Quantum Network, University of Science and Technology of China, Hefei 230026, China}

\author{Chang-Ling~Zou}
\email{clzou321@ustc.edu.cn}
\affiliation{Laboratory of Quantum Information, University of Science and Technology of China, Hefei 230026, China.}
\affiliation{Anhui Province Key Laboratory of Quantum Network, University of Science and Technology of China, Hefei 230026, China}
\affiliation{Hefei National Laboratory, Hefei 230088, China}

\author{Baile~Chen}
\email{chenbl@shanghaitech.edu.cn}
\affiliation{School of Information Science and Technology, ShanghaiTech University, Shanghai, China}

\author{Luyan~Sun}
\email{luyansun@tsinghua.edu.cn}
\affiliation{Center for Quantum Information, Institute for Interdisciplinary Information
Sciences, Tsinghua University, Beijing 100084, China}
\affiliation{Hefei National Laboratory, Hefei 230088, China}

%\date{\today}

\begin{abstract}
\textbf{
Superconducting quantum computing has the potential to revolutionize computational capabilities. However, scaling up large quantum processors is limited by the cumbersome and heat-conductive electronic cables that connect room-temperature control electronics to quantum processors, leading to significant signal attenuation. Optical fibers provide a promising solution, but their use has been restricted to controlling simple two-level quantum systems over short distances. Here, we demonstrate optical control of a bosonic quantum processor, achieving universal operations on the joint Hilbert space of a transmon qubit and a storage cavity. Using an array of cryogenic fiber-integrated uni-traveling-carrier photodiodes, we prepare Fock states containing up to ten photons. Additionally, remote control of bosonic modes over a transmission distance of 15\,km has been achieved, with fidelities exceeding $95\%$. The combination of high-dimensional quantum control, multi-channel operation, and long-distance transmission addresses the key requirements for scaling superconducting quantum computers and enables architectures for distributed quantum data centers.
%cooperative control of a transmon qubit and cavities in a bosonic quantum processor via optical links, employing an array of fiber-integrated uni-traveling-carrier photodiode chiplets. We achieve high-fidelity control of composite qubit and bosonic mode system by preparing Fock states up to ten photons, and demonstrate key aspects of scalability of optical links, including the synchronized multi-channel operations, universal control of high-dimension Hilbert space, and the remote control over a transmission distance of 15\,km. Our results pave the way for large-scale quantum computing where high-density control lines are critical, and also provide essential insights for quantum data centers and distributed quantum computation.
%high precision in preparing large, high-dimensional Fock states, a far more challenging task than qubit control. Additionally, we extend the optical links to 15\,km, with negligible fidelity degradation, illustrating two key aspects of scalability: (1) the ability to control large Hilbert space quantum systems and (2) the potential for scaling up quantum processors across data centers or long-distance quantum communication networks. Our results pave the way for large-scale, fault-tolerant quantum computing and provide essential insights for the next generation of quantum technologies, where low-latency operations are critical for quantum networking and distributed quantum computation.
}
\end{abstract}

\maketitle

\noindent \textbf{Introduction}{\large\par}
\noindent Superconducting quantum processors have emerged as a leading platform for quantum computation, offering fast gate operations on the order of tens of nanoseconds, fidelities exceeding 99.9\% for single-qubit gates, and natural compatibility with integrated circuit fabrication ~\cite{PhysRevLett.134.090601,RN1,RN4}. Recent milestones have demonstrated quantum error correction (QEC) below the surface code threshold~\cite{RN2} and scalable logical operations in color codes on multi-qubit arrays~\cite{RN3}. In parallel, bosonic quantum modules that encode logical information in the infinite-dimensional Hilbert space of superconducting cavities have achieved logical qubit lifetimes surpassing those of their constituent physical components, establishing hardware-efficient pathways toward fault tolerance~\cite{RN5,RN6,Cai2021,RN7,RN25}. These advances collectively validate the viability of the superconducting approach and motivate continued efforts to scale these systems.

However, a critical engineering challenge threatens to limit further progress: the ``wiring bottleneck"~\cite{Mukai_2020,RN8}. Current superconducting quantum processors require individual microwave control lines for each qubit, with signals generated by room-temperature electronics and transmitted through coaxial cables to processors operating at millikelvin temperatures~\cite{Krantz_2019}. This architecture introduces three compounding problems~\cite{Krinner_2019,EPJQT9,RN10}. First, the metallic cables conduct heat directly into the cryogenic environment, with thermal dissipation scaling linearly with the number of control channels. Second, microwave signals suffer substantial attenuation over the meter-scale distances between room-temperature electronics and the mixing chamber, necessitating attenuators at each temperature stage that further increase heat load. Third, the space requirement of coaxial cables limits channel density, constraining the number of qubits that can be accommodated within a single dilution refrigerator. Conservative estimates suggest these factors will restrict conventionally-wired systems to several thousand qubits~\cite{joshi2022scalingsuperconductingquantumcomputers}, far below the millions required for fault-tolerant operation~\cite{Ahsan_2015,RN11,PhysRevA.86.032324}.

Several approaches have been proposed to address this challenge. Cryogenic CMOS control electronics can reduce cable count by moving signal generation closer to the quantum processor~\cite{RN8,PRXQuantum.5.010326,RN12,RN13,RN14}, though heat dissipation from active electronics remains. Frequency multiplexing allows multiple qubits to share control lines ~\cite{shi2023multiple,AIPMultiplexed,Chapman_2017} but introduces crosstalk and constrains circuit design~\cite{wang2022control,PRXQuantum.3.020301}. Optical approaches, including qubit readout and optical interconnects for distributed quantum computing, have been demonstrated recently~\cite{RN16,RN18,Li24,van_Thiel_2025,RN19,RN20,zhou2025kilometer}. However, substantial gaps remain in existing optical methods as previous demonstrations have been restricted to single-channel operation on simple two-level qubit systems, with a meter-scale transmission distances. The simultaneous achievement of high-dimensional quantum control, long-distance transmission, and multi-channel scalability has not been demonstrated yet.

\begin{figure}
\begin{centering}
\includegraphics[width=1\linewidth]{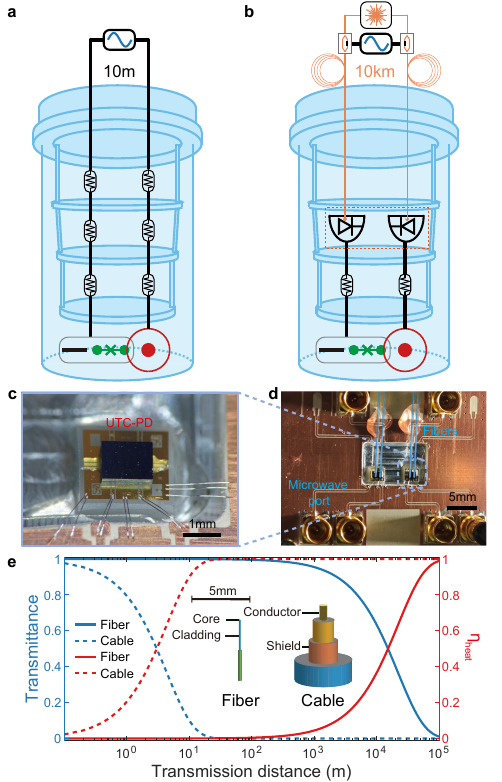}
\par\end{centering}
\caption{\textbf{Optical links for controlling superconducting quantum processor units}. \textbf{a} and \textbf{b}, Comparison of conventional microwave link (left) and optical link (right) for delivering control signals from room temperature electronics facilities to superconducting devices in a dilution refrigerator. Microwave links employ coaxial cables spanning multiple temperature stages with attenuators at each stage, while optical links transmit modulated optical carrier through optical fibers and recover the microwave signal through a UTC-PD.  \textbf{c}, Photograph of a UTC-PD chiplet, which contains 6 elements and two of which are wire-bonded to a printed circuit board for microwave output. \textbf{d}, Packaged UTC-PD array with two separate chiplets, showing glued optical fibers to 4 elements. \textbf{e}, Dependence of transmittance and the ratio of carrier power dissipation $\eta_\mathrm{heat}$ on the transmission distance in coaxial cables (dashed line) and optical fibers (solid line). Inset: the structures of an optical fiber and a coaxial cable. }

\label{Fig1}
\end{figure}
Here, we introduce a scalable optical link approach for quantum control based on custom-fabricated fiber-integrated chiplets of uni-traveling-carrier photodiode (UTC-PD) array, chosen for their high bandwidth and linearity~\cite{RN22,Li10475411,Song:21}, that overcomes these limitations. We demonstrate three critical dimensions of scalability that collectively establish the viability of optical control links for large-scale superconducting quantum processors.  First, we achieve precise control over high-dimensional Hilbert spaces, demonstrated by the preparation of Fock states up to $10$ microwave photons in a bosonic cavity using Gradient Ascent Pulse Engineering (GRAPE) optimized pulses~\cite{KHANEJA2005296}. This capability extends far beyond simple two-level qubit manipulation and validates the compatibility of optical control with bosonic QEC schemes that encode logical qubits in cavity modes. Second, we extend the optical link to $15~\mathrm{km}$ of optical fiber while maintaining high gate fidelity, demonstrating that control electronics can be physically decoupled from the quantum processor, without compromising control precision. Third, we implement a $2\times2$ multi-channel UTC-PD array that simultaneously delivers independent control signals to multiple quantum devices, paving the way for scaling to large qubit arrays. The combination of these three scalability dimensions, achieved within a single integrated platform, provides a viable pathway toward fault-tolerant quantum computers with millions of qubits and distributed quantum computation architectures requiring dense, low-latency, and high-fidelity control channels over extended distances.

\smallskip{}
\noindent \textbf{\large{}{}Results}{\large\par}

\noindent Figure~\ref{Fig1} schematically illustrates and compares the microwave and optical links approaches for controlling quantum computing systems in cryostats. In the conventional approach (Fig.~\ref{Fig1}a), microwave signals are transmitted via bulky coaxial cables that span from room-temperature electronics to the milli-Kelvin stage. This configuration creates a direct thermal path, introducing heat load through both thermal conduction of the metal and the necessary dissipation from attenuators at each cooling stage~\cite{Simbierowicz2024}. The optical approach (Fig.~\ref{Fig1}b) replaces these metallic links with optical fibers. By encoding microwave control signals onto an optical carrier at room temperature, we leverage the fiber's negligible thermal conductivity and low loss to deliver signals directly to the cryogenic environment. The optical-to-microwave conversion occurs locally at the 4K stage using photodiodes, thereby physically decoupling the signal generation from the quantum processor and eliminating the cascading thermal anchors required by coaxial chains.

\begin{figure*}[t]
\begin{centering}
\includegraphics[width=1\linewidth]{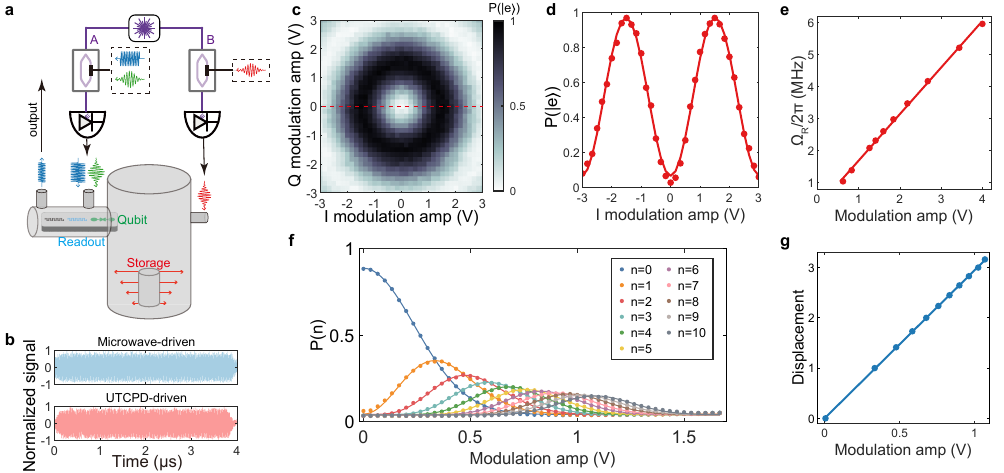}
\par\end{centering}
\caption{\textbf{Independent optical control of components in a bosonic quantum processor.} \textbf{a}, Schematic of the optical links (A and B) and the bosonic quantum processor comprising a transmon qubit, a readout resonator, and a high-quality storage cavity. \textbf{b}, Transmitted microwave signals of the readout resonator with input from microwave link (blue) and optical link (red). Input pulses are $4~\mathrm{\mu s}$ square waves at $7.508~\mathrm{GHz}$, are down-converted to $50~\mathrm{MHz}$ for pulse shape acquisition. \textbf{c}, The excitation probability $\text{P}(\ket{e})$ of the transmon qubit driven by optical link A, using Gaussian pulse modulation on the electro-optics modulator with pulse-width $\sigma=60~\mathrm{ns}$ and complex modulation amplitude in the IQ plane.
\textbf{d}, Line cut along the red dashed line in \textbf{c}, showing qubit Rabi oscillations as a function of the I modulation amplitude. \textbf{e}, Linear dependence of the qubit Rabi frequency ($\Omega_\text{R}$) on the modulation amplitude. \textbf{f}, Cavity microwave photon number distribution as a function of the modulation amplitude, with the curves are fittings according to Poisson distribution for coherent states with different displacement amplitude. \textbf{g}, Linear relationship between the coherent state displacement amplitude and the modulation amplitude.}
\label{Fig2}
\end{figure*}

Figures~\ref{Fig1}c and \ref{Fig1}d display the compact UTC-PD chiplets, which are central to our optical link. The chiplets coherently convert modulated optical signals into microwave signals, preserving the phase coherence while losing the quantum superposition properties of the input which is distinct from the quantum transducers for microwave-optical frequency conversion~\cite{van_Thiel_2025,RN19,RN20}. We fabricate an array of UTC-PD elements on a single chiplet through a scalable lithographic process, and then package it with precisely aligned optical fiber pigtails, establishing multiple independent control channels within a compact footprint. Unlike active cryogenic electronics, these UTC-PDs operate as passive devices without external bias power, ensuring that heat generation is limited strictly to the absorbed optical power and photocurrent dissipation. The chiplet-based architecture also supports a hierarchical integration strategy: individual chiplets can be characterized and pre-tested at room temperature before cryogenic integration for high fabrication yield.

The advantage of optical fiber over coaxial cable is quantitatively studied in Fig.~\ref{Fig1}e. Optical fibers exhibit attenuation ($0.2~\mathrm{dB/km}$~\cite{Li2020}) nearly four orders of magnitude lower than coaxial cables ($\sim 1000~\mathrm{dB/km}$ at $6~\mathrm{GHz}$), thereby enabling kilometer-scale transmission with negligible signal loss and a corresponding reduction in  heating from absorption. Furthermore, the radius of a standard coaxial cable is about 50 times larger than that of an optical fiber, offering substantial advantages in space management for controlling large arrays of superconducting qubits and bosonic modes.

To validate the optical link's capability in controlling a high-dimensional quantum system, we integrate it with a 3D superconducting module~\cite{PhysRevLett.107.240501,PhysRevB.94.014506,RevModPhys.93.025005} comprising a readout resonator, a transmon qubit, and a storage cavity, as illustrated in Fig.~\ref{Fig2}a. The transmon qubit is dispersively coupled to both the high-quality storage cavity and the readout resonator, with the system dynamics governed by the Hamiltonian
\begin{equation}
\begin{aligned}
H=&\omega_\mathrm{r} \hat{a}_\mathrm{r}^\dagger\hat{a}_\mathrm{r}+\omega_\mathrm{s} \hat{a}_\mathrm{s}^\dagger\hat{a}_\mathrm{s}+\omega_\mathrm{q} \ket{e}\bra{e}\\&-\chi_\mathrm{rq} \ket{e}\bra{e} \hat{a}_\mathrm{r}^\dagger\hat{a}_\mathrm{r}-\chi_\mathrm{sq} \ket{e}\bra{e} \hat{a}_\mathrm{s}^\dagger\hat{a}_\mathrm{s}-\frac{K_\mathrm{g}}{2}\hat{a}_\mathrm{s}^\dagger\hat{a}_\mathrm{s}^\dagger\hat{a}_\mathrm{s}\hat{a}_\mathrm{s},
\end{aligned}
\label{Eq1:systemH}
\end{equation}
where $\omega_\mathrm{r}$  and $\omega_\mathrm{s}$ are the frequencies of the readout resonator $\hat{a}_\mathrm{r}$ and the storage cavity mode $\hat{a}_\mathrm{s}$, respectively, $\omega_\mathrm{q}$ is the transition frequency of the transmon qubit between the ground state ($\ket{g}$) and the first excited state ($\ket{e}$), $\chi_{\text{rq}}/{2\pi}=1.20\,\mathrm{MHz}$ $(\chi_\mathrm{sq}/{2\pi}=0.417\,\mathrm{MHz})$ is the cross-Kerr interaction between the readout resonator (the storage cavity) and the transmon, and $K_\text{g}/{2\pi}=1.32\,\mathrm{kHz}$ is the self-Kerr of the storage cavity. The measured frequencies for the qubit, storage cavity, and readout resonator are $\omega_\mathrm{q}/2\pi = 4.257$\,GHz, $\omega_\mathrm{s}/2\pi = 5.922$\,GHz, and $\omega_\mathrm{r}/2\pi = 7.508$\,GHz, respectively. The structure and the fiber-link control of the 3D superconducting bosonic quantum processor are shown in Fig.~\ref{Fig2}a. The readout resonator and the qubit share a common input port, while the storage cavity uses a separate port. Consequently, two separate optical links are introduced to control the system. Each link employs an electro-optic modulator (EOM) at room temperature to encode microwave waveforms onto an optical carrier, with the modulation frequency matched to the target mode. The modulated light propagates through optical fibers to the cryogenic UTC-PD array, which converts the signals back to microwave frequencies for driving the quantum system.

We first characterize the optical link A for the transmon qubit control and readout. {The qubit state is determined via the dispersive coupling between the qubit and readout resonator, as the transmitted  microwave signal acquires distinct phases and amplitudes for qubit in the ground and excited states~\cite{PhysRevA.69.062320,RevModPhys.93.025005}. These differences are detected by conventional room-temperature electronics after the signal exits the dilution refrigerator through a separate output line.} As shown in Fig.~\ref{Fig2}b, the transmitted signals generated by the UTC-PD through the optical link faithfully  reproduce the waveforms from conventional microwave driving, confirming that the optical link preserves the pulse shape without significant distortion.
For qubit control, Fig.~\ref{Fig2}c plots the qubit excitation probability under varying modulation amplitude, demonstrating optical control along arbitrary directions in the IQ plane. We observe clear Rabi oscillations when driving the transmon via the optical link. The concentric patterns in the IQ plane indicate uniform driving strength along all phase directions, while the cut in Fig.~\ref{Fig2}d exhibits sinusoidal Rabi oscillations, indicating that the qubit drive operates within the linear response regime. Furthermore, the  Rabi frequency of qubit scales linearly with the modulation amplitude ({Fig.~\ref{Fig2}e}). This confirms that the optical link driving the qubit transition operates in the linear regime, which is convenient for precise gate calibration.

The optical link B is introduced for controlling the bosonic mode, and Fig.~\ref{Fig2}f presents results of the simplest displacement operation  from vacuum to prepare a coherent state with varying amplitudes. {The cavity microwave photon number distribution is measured by mapping the population of each Fock state $\ket{n}$ to the qubit via photon-number-selective $\pi$-pulses, followed by qubit state readout.} %The cavity microwave photon number distribution is measured by preparing the qubit in the $\ket{g}$ state and mapping the population of each Fock state $\ket{n}$ onto the qubit, such that the probability of finding the qubit in $\ket{e}$ reflects the Fock state occupation. This is done via a series of photon-number-selective $\pi$ pulses, followed by qubit state readout.}
The measured distribution closely follows the expected Poisson distribution, $P(n) = e^{-\bar{n}} {\bar{n}^n}/{n!}$, with $\bar{n}$ denoting the mean photon number. The linear relationship between the cavity displacement amplitude and the modulation amplitude, shown in Fig.~\ref{Fig2}g, confirms the linearity of the optical link for cavity control. Throughout these measurements, we observe no measurable crosstalk between channels, confirming the frequency selectivity required for independent and simultaneous control of multiple devices.

\begin{figure}[t]
\begin{centering}
\includegraphics[width=1\linewidth]{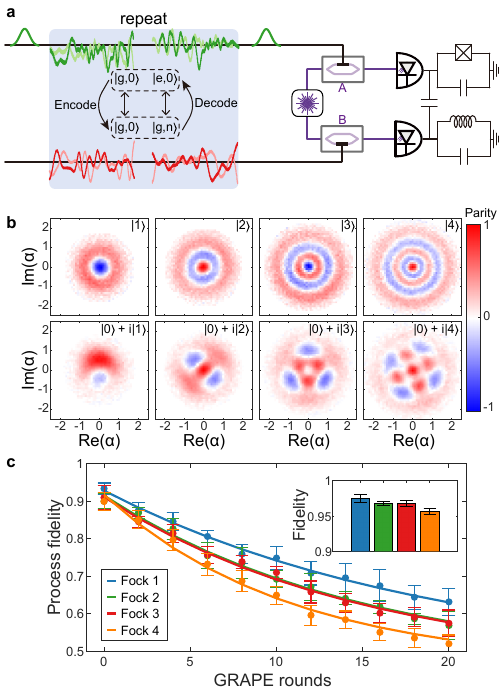}
\par\end{centering}
\caption{\textbf{Quantum control of the joint cavity-qubit system via optical links.} \textbf{a}, Schematic of arbitrary operations on the bosonic quantum processor through two optical links. Inset: the synchronized pulse sequences on both optical links A and B for implementing encoding and decoding operations between qubit states \{$\ket{g}$, $\ket{e}$\} and cavity Fock states \{$\ket{0}$,$\ket{n}$\}. \textbf{b}, Gallery of measured Wigner functions for prepared cavity states. Top row: Fock states $\ket{n}$ for $n=1$ to $4$. Bottom row: superposition states $\ket{0}+i\ket{n}$. The corresponding state fidelities are $96.9\%(1),90.1\%(2),97.7\%(3),88.3\%(4)$ and $95.0\%(1),90.4\%(2),92.1\%(3),85.6\%(4)$, respectively. \textbf{c}, Process fidelity against the rounds of encoding and decoding operations for Fock states with $n=1$ to $4$. Inset: estimated encoding operation fidelity.}
\label{Fig3}
\end{figure}

Having verified the basic functionality of the two optical links for delivering coherent signals to the bosonic mode and the transmon qubit, we now examine their performance in more challenging tasks for controlling highly non-classical states in the infinitely large Hilbert space of a bosonic mode. In contrast to driving a bosonic mode or executing single-qubit gates in two-dimensional Hilbert space, which requires only simple pulse shapes at fixed frequencies, the universal control of a high-dimensional bosonic system demands precise, synchronized, and complex pulse sequences that act simultaneously on the dispersively coupled transmon and storage cavity~\cite{RevModPhys.93.025005}. This level of control has not yet been demonstrated through optical links. Building upon the validated independent control, we advance to the universal control over the joint Hilbert space of the qubit and the storage cavity via GRAPE-optimized pulses~\cite{Heeres2017,KHANEJA2005296,Krastanov2015, Reinier2015,Eickbusch2022}, which generate optimized pulse sequences with intricate, time-varying amplitude and phase profiles tailored to the system Hamiltonian.

Figure~\ref{Fig3}a illustrates the experimental setup for the universal control of joint cavity-qubit system through two optical links simultaneously. As an example of universal control, we design the GRAPE pulses for an encoding operation, which maps the qubit state $\ket{g}$ and $\ket{e}$ to the cavity Fock states $\ket{0}$ and $\ket{n}$, respectively. It is worth noting that such an operation manifests a non-trivial qubit-cavity swap gate that technically is more challenging than single-qubit operations realized in Ref.~\cite{RN16,RN18,Li24}. In Fig.~\ref{Fig3}b, we show a gallery of Wigner functions $W(\alpha)$ of the prepared cavity states. The Wigner function offers a complete phase-space representation of the bosonic quantum state, where negative values are a hallmark of quantum behavior. We successfully prepare Fock states from $\ket{n=1}$ through $\ket{n=4}$, as evidenced by the concentric ring structures that are insensitive to the phase degree-of-freedom. As a comparison, the superposition states $(\ket{0}+i\ket{n})/\sqrt{2}$ exhibit $n$-fold rotational symmetry. The distinct negativity observed in these Wigner functions confirms that the optical links can implement complex control pulses on the composite system in an extended Hilbert space.

\begin{figure*}[t]
\begin{centering}
\includegraphics[width=1\linewidth]{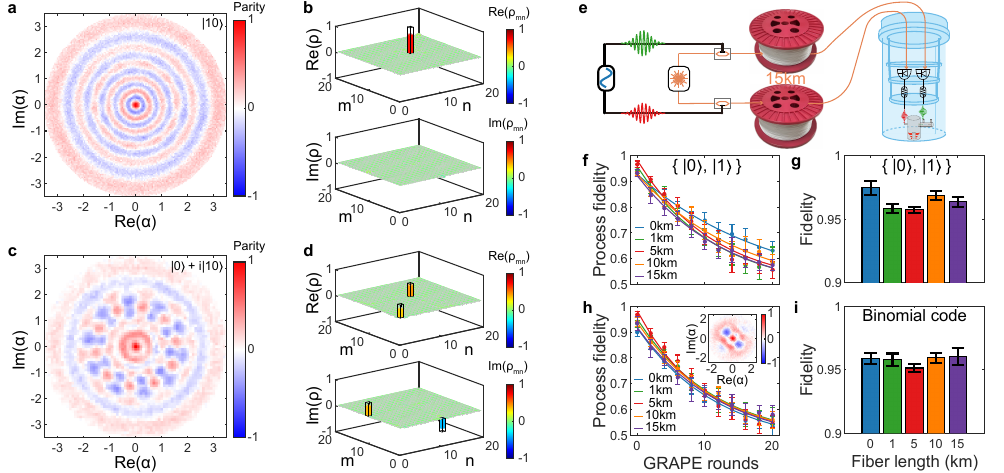}
\par\end{centering}
\caption{
\textbf{Scalability of optical links in Hilbert space dimension and transmission distance.} \textbf{a}-\textbf{d}, Measured Wigner functions and the reconstructed density matrices of the Fock state $\ket{10}$ and its superposition state $\ket{0} + i\ket{10}$, respectively. \textbf{e}, Schematic of remote control of bosonic quantum processor via optical link over a long transmission distance. \textbf{f} and \textbf{g}, Process fidelity decay and extracted fidelity of encoding operation for Fock state basis \{$\ket{0}$, $\ket{1}$\}, with varying fiber spools lengths ($0-15\,\mathrm{km}$) inserted between room-temperature electronics and the cryogenic processor. \textbf{h} and \textbf{i}, Process fidelity characterizations for binomial code logical state basis \{$\ket{0_L}=(\ket{0}+\ket{4})/\sqrt{2}$, $\ket{1_L}=\ket{2}$\}. The inset in \textbf{h} shows the results of the Wigner tomography of the state $\ket{0_L}+i\ket{1_L}$ at $0\,\mathrm{km}$.}
\label{Fig4}
\end{figure*}

Due to the higher dimensions involved in preparing states with larger Fock states $n$ and the readout errors, the fidelities of the prepared states degrade with $n$. To quantitatively evaluate the control fidelity of the GRAPE-based operations, we employ a benchmarking protocol based on repeated encoding and decoding operations, as illustrated in Fig.~\ref{Fig3}c. Here, each decoding operation maps the cavity Fock states $\ket{0}$ and $\ket{n}$ back to the qubit states $\ket{g}$ and $\ket{e}$, respectively. By applying many rounds of encoding and decoding operations, we extract the process fidelity of one round of operations through the reconstructed process matrix of final qubit state. For Fock states from $\ket{n=1}$ to $\ket{n=4}$, the process fidelity begins above 0.9 and decreases gradually with increasing operation rounds. The decay reflects the accumulation of errors from finite coherence times and residual control imperfections, but the high initial fidelities confirm that the optical links support precision quantum control. Notably, encoding and decoding of lower Fock states maintain higher fidelity at each round, consistent with their reduced sensitivity to errors in smaller Hilbert spaces. The inset in Fig.~\ref{Fig3}c shows the operation fidelities for single encoding gates (average fidelity derived from one round operation fidelity), all exceeding $95\%$.
These results demonstrate that the photonic control link supports complex quantum operations with fidelities comparable to conventional microwave delivery, imposing no fundamental limitations on operation complexity.

Having established high-fidelity control, we investigate two critical dimensions of scalability essential for practical superconducting quantum computing: the capacity to access large Hilbert spaces for bosonic QEC~\cite{Cai2021} and the ability to deploy remote control over extended distances for distributed quantum computing~\cite{Cirac1999}. To demonstrate the scalability in Hilbert space dimension, we employ a $6\,\mu\text{s}$ GRAPE-optimized pulse to prepare high-photon-number state $\ket{10}$ and its superposition $\ket{0}+i\ket{10}$ through optical links. The resulting Wigner function for $\ket{10}$ (Fig.~\ref{Fig4}a) clearly resolves 10 concentric rings, while the superposition state (Fig.~\ref{Fig4}c) displays the expected ten-fold rotational symmetry. Density matrix reconstruction via maximum likelihood estimation~\cite{10Tomography} reveals state fidelities of $75.9\%$ and $78.2\%$, respectively.

The scalability in transmission distance is investigated by inserting fiber spools ranging from 0\,km to 15\,km between the room-temperature control electronics and the cryogenic processor, as depicted in Fig.~\ref{Fig4}e. We repeat the encoding and decoding operations for the lowest Fock state encoding on \{$\ket{0}$, $\ket{1}$\} for different transmission distances, and reveal similar fidelity decay against operation rounds in Fig.~\ref{Fig4}f, with the corresponding fidelities of single encoding operations extracted and shown in Fig.~\ref{Fig4}g. We further extend the encoding and decoding to higher dimensions employing the binomial QEC codes, i.e., realizing the mapping between the logical qubit basis \{$(\ket{0}+\ket{4})/{\sqrt{2}}$, $\ket{2}$\} and the qubit states \{$\ket{g}$, $\ket{e}$\}. The results are summarized in Figs.~\ref{Fig4}h and \ref{Fig4}i. Remarkably, the operation fidelities consistently exceed 95\% regardless of the transmission distances. This insensitivity to fiber length reflects a fundamental advantage of the optical links, confirming that the control electronics can be physically decoupled from the quantum processor over kilometer-scale distances without compromising operation precision.

\smallskip{}
\noindent \textbf{\large{}{}Discussion}{\large\par}
\noindent We have demonstrated quantum control of a bosonic quantum processor through optical links comprising optical fibers and cryogenic UTC-PD chiplets. The successful execution of GRAPE-optimized pulses for Fock state preparation up to ten microwave photons, in a composite transmon-qubit and bosonic-mode system, establishes that optical links support universal, high-precision operations on high-dimensional quantum systems, a capability far beyond simple two-level qubit manipulation.

We highlight the scalability of optical links for realizing large-scale quantum computation from three aspects. First, the ultralow attenuation of optical fibers enables long-distance remote control of quantum processors. For modular designs of large-scale quantum computers, optical links allow a centralized classical control system to distribute synchronized signals across multiple quantum processors while mediating their classical communication with minimal latency. Second, the negligible thermal conductivity and compact footprint of optical fibers dramatically reduce both the thermal budget and space requirements in the cryogenic environment, relaxing constraints on the number of qubits or devices that can operate within a single dilution refrigerator. Third, the cost of optical links and supporting infrastructure is low. Equipment and devices operating at telecom wavelengths are mature and widely deployed in conventional communication systems, enabling industrial-scale manufacturing and deployment.

Our results demonstrate that optical links impose no fundamental limitations on quantum operation complexity or fidelity. Although the current implementation is restricted to one or two control channels per fiber, we anticipate that channel capacity can be substantially increased by incorporating optical microcomb sources~\cite{chang2022integrated}, wavelength-division multiplexing components~\cite{dong2016silicon}, and on-chip electro-optic modulators~\cite{wang2018integrated} integrated with UTC-PD chiplets via flip-chip assembly. Furthermore, this optical-link approach can be extended to other cryogenic electronic control applications, including semiconductor quantum processors.

\smallskip{}

\noindent \textbf{\large{}{}Acknowledgment}{\large\par}

\noindent This work was funded by the Quantum Science and Technology-National Science and Technology Major Project (Grant Nos.~2024ZD0301500 and 2021ZD0300200), the National Natural Science Foundation of China (Grant Nos.~92265210, 123B2068, 92165209, 92365301, 12474498, 12374361, and 12293053). We also acknowledge the support from the Fundamental Research Funds for the Central Universities and USTC Research Funds of the Double First-Class Initiative. The UTC-PD chips were fabricated with support from the ShanghaiTech University Material and Device Lab (SMDL). The numerical calculations in this paper were performed on the supercomputing system in the Supercomputing Center of USTC, and this work was partially carried out at the USTC Center for Micro and Nanoscale Research and Fabrication.

\end{document}